# AskDoc - Identifying Hidden Healthcare Disparities

Shashank Gupta
Department of Computer Science
University of Kentucky
shashank.gupta@uky.edu

## Abstract

**Objectives**: To study the online Ask the Doctor services medical advice on internet platforms via AskDoc, a Reddit community that serves as a public AtD platform, and study if platforms mirror existing hurdles and partiality in healthcare across various demographic groups.

**Methods**: We downloaded data from January 2020 to May 2022 from the AskDoc, a subreddit, and created regular expressions to identify self-reported demographics (Gender, Race, and Age) from the post, and performed statistical analysis to understand the interaction between peers and physicians with the posters.

**Results**: Half of the posts didn't receive comments from peers or physicians. At least 90% of the people disclose their gender and age, and 80% of the people do not disclose their race. It was observed that r/AskDocs is dominated by adult (age group 20-39) white males. Some disparities were observed in the engagement between the users and the posters with certain demographics.

**Conclusion**. Beyond the confines of clinics and hospitals, social media could bring patients and providers closer together however as observed current physicians' participation is low as compared to posters.

## Abbreviations and Acronyms

**AtD**: Ask the Doctor, **RegEx**: Regular Expression, **e.g.:** For example, **PRAW**: Python Reddit API Wrapper, **PSAW**: Python Pushshift.io API Wrapper, **OP**: Original Poster, **WW**: White Woman, **AA**: American African, **id**: Identity

## Introduction

Recent years have seen a significant revolution in the usage of the internet for social media, allowing for human connection and engagement in ways that go beyond conventional face-to-face encounters. This increased usage of social media has been widely observed during the recent coronavirus disease outbreak. According to a study by Kantar, a London-based data and insights company, during the initial wave of the pandemic, social media involvement soared by 61%.[10] Online health consultation via web platforms increased 78 times from February 2020 to April 2020 early in the COVID-19 pandemic as patients and providers sought safe methods to access and deliver healthcare while maintaining social distancing.[22] Apart from using social media for entertainment, e-commerce, and various other reasons, social media has also been widely employed by professionals and the public in healthcare to bridge communication gaps, creating new techniques and chances for connection. The present Ask the Doctor (AtD) service is frequently text-based where patients submit health-related queries online, which are subsequently addressed by healthcare practitioners enrolled on the platform. According to a survey, an average American waits 24 days to get a new physician appointment.[5] People have opted to use the internet to search for healthcare advice because of its less cost, more comfort, and continuous day-and-night availability. Inequalities between demographic groups in health and medical care are a long-standing and chronic problem in the US. Healthy People 2020 defines "*health equity*" as the "*attainment of the highest level of health for all people*" and defines a "*health disparity*" as "*a particular type of health difference that is closely linked with social, economic, and/or environmental disadvantage*.[13] There is a discrepancy when a health consequence is seen to a greater or lower level in different groups. Race or ethnicity, gender, sexual identity, age, disability, financial level, and geographic location all have an impact on an individual's potential to achieve optimal health.[13] Racial/ethnic and gender inequities in healthcare are noteworthy as they provide serious moral and ethical quandaries for the US healthcare system. Here are a few instances to illustrate: Infant mortality for black babies remains nearly 2.5 times higher than for white babies; life expectancy for black men and women remains nearly a decade lower than for white counterparts; diabetes rates are more than 30 percent higher among Native Americans and Latinos than among whites; rates of death from heart disease, stroke, and prostate and breast cancers remain much higher in black populations; and minorities remain grossly underserved.[1] As a result, many online platforms emerged to mitigate the bias in healthcare delivery. For example, WebDoctors[21], an online paid telemedicine portal, where patients provide a piece of basic medical information and board-certified physicians make diagnosis for a charge. This research investigates how people are utilizing free social media to seek medical advice, and if there is any bias experienced among various demographic groups.

As one of the most popular social platforms on the web, Reddit is the 8th most-visited website in the world and the 7th most-visited website in the United States as of June 2022, according to Semrush.[12] The most traffic on Reddit is generated by United States (45.28%) followed by Canada (7.09%), United Kingdom (5.62%), Germany (4.06%) and Australia (2.76%). Reddit has 430 million monthly active users and over 3.1 million topical communities called "subreddits" (only 138,000 are currently active).[11] Subreddits are user-created forums where discussions related to subjects are organized. With more than 395k members (as of June 2022), r/AskDocs is a health Q&A subreddit that was created in July 2013 that allows individuals to seek medical advice, which is addressed by other users (layperson) and physicians which are verified by the subreddit moderator (can be seen by the tags as Clinical Psychologist, Physician, etc.). It offers the benefit of being free of charge, continuous day-and-night availability all being anonymous by using throw-away accounts. Racial and sex minorities frequently have poorer access to quality healthcare than their counterparts, regardless of insurance coverage, income, age, or severity of diseases. Even when they have access, minorities usually receive inferior quality treatment and have poorer health outcomes than their peers for identical health concerns.[3][2]

We analyze the following topics in this paper:

1. **What is the number of self-reported demographics (gender/sex, Race/Ethnicity, and age)?** This analysis could answer a few questions such as why some people report demographics and what are the reasons others don't. Do people from certain demographics feel that disclosing their demographics could lead to less or more responses from physicians and other users and would not get proper health advice?

2. **Analysis of the comments by users and physicians received on the posts of people with various demographics**. This analysis is done to study the response by the physicians and other users that how they engage with posters with various demographic groups. This study could determine if users' or physicians' response to the post depends on the posters' demographics?

3. **Posters Engagement with the users and the physicians on the AskDoc subreddit**. Engagement is defined as an OP's interaction or replies to at least one of the non-OP comments on the post. This analysis could help us to observe how people with certain demographics interact with other users who responded to their posts. It could help us to study the relationship between the poster's engagement and the number of responses received. For example, if the poster does not engage in the discussion, it is likely to receive less response from other users as compared to the poster who engages more.

Similar work has been done by Nobles et al.[2] for the period (July 2013 to December 2018) where they performed an analysis on self-reported demographics for gender and race on Reddit. Our analysis extends this work by including age in the demographics. Secondly, in addition to male, female, and transgender, we study the interaction of non-binary posters as well. Also, we performed a more fine-grained analysis of the interaction of the posters and the users through the comments. Understanding how online health care differs from conventional healthcare professional-patient interactions will be crucial to providing good quality treatment as the industry expands.

## Data Collection

Moderator wants posters to provide as much as information as possible. This is the requirement of submission: "Please include age and sex (write as [age][sex], e.g., 18M), height, weight, race, primary complaint, duration, any existing medical issues, current medications, and doses, and whether you drink, smoke, and/or use recreational drugs. For all visible issues, particularly dermatological, a photo is not required, but helpful." But not everyone disclosed their demographics hence many posts do not contain any/some of the demographics.

From January 2020 through May 2022, we gathered all r/AskDocs posts, comments, and related information using PRAW (Python Reddit API Wrapper) and PSAW (Python Pushshift.io API Wrapper) which requires setting up an authorization token, so that Reddit knows about the application. Since Reddit allows throw-away accounts (a throwaway account is a temporary account that is not a primary Reddit account), many of the users deleted their account after posting or commenting about their health issues due to privacy reasons. The username of such users become unavailable and hence was renamed to "none". Hence "none" is also one of the users in our study. Every post contains an automatic (bot) response from the subreddit moderator; hence this comment was removed from our study. Our final dataset (Table 1) included 177,850 posts from 121,485 unique posters where 89050 (50.08%) of posts contained at least one comment from a user other than a poster (non-OP). The rest of the posts either contain no comments 86450 (48.6%) or comments from poster only 2350 (1.32%). These posts contain 426,854 comments from 85,423 unique commenters. Physicians contributed 127013 (29.8%) of all comments which were from 1641 unique physicians.

## Method

We extracted self-reported Demographics in posts. Though r/AskDocs requests demographic information such as gender, race, and age, no specific format was seen. For example, "I'm 24F, nonsmoker, only drink on special occasions, no recreational drugs" identifies a 24-year-old female with no disclosure of her race. Another post contained "*22 year old male, Caucasian, 6 feet, 177lb*." which identifies a 22-year-old male of the white race. Texts were pre-processed to lowercase and some of the punctuations were removed. People interchangeably used ethnicity with race (although Hispanic/Latina are ethnicity, posters frequently use it as a race). The final categories for races are White, Black, Asian, Hispanic/Latina, American Indian, and Hawaiian. Similarly, we identified self-reported genders/sex, and the final categories are Male, Female, Transgender, and Non-Binary. Similarly, self-reported age was identified and was categorized as per the range in which age falls. The final categories for age are Children (1-12), Teenagers (13-19), Adults (20-39), Middle Age (40-59), and seniors (60+). Posts that do not disclose gender, race, or age were categorized as Unknown.

### 1. Gender Extraction

We first created some regular expressions to extract the gender (male, female, non-binary, and trans) mentioned in the title or the body of the post. We used patterns like (male/guy/man etc.) for identifying males, (woman/female, etc.) to identify female, (agender/bigender/demi gender, etc.) for non-binary and (transgender/MtF/F2M, etc.) to identify trans genders respectively. If gender is not recognized, it is categorized as unknown.

## 2. Race Extraction

By looking at a few hundred posts, we observe that people disclose their demographic details within the vicinity of 8-10 tokens. For Example, posts like "33/F black 150 lbs. I've got a small lump on the right side of my neck and on the left cheek of my face. There is a white discharge from the lump." Here actual race is black whereas the white adjective is used for the lump discharge. Similarly, people can talk about black toenails, Asian food, etc. Hence, we cannot directly use regex patterns like (white/black/Asian, etc.) on the entire post to identify race as the tokens do not necessarily mean the sentence contains the race. Hence, we assume that people disclose their race around/near their gender.

To extract race, we formed the extracted gender token/tokens from the step mentioned in part 1 as the center word, and we created a window of 10 unigrams and bigrams, both to the left and right of the center word in the post. We then search for race patterns like (Caucasian/white/ww, etc.) to identify white, (black/African American/aa, etc.) for blacks, (Latin[o, a, x]/Hispanic, etc.) for Hispanic/Latina, (Asian) for Asians, (native American/American Indian, etc.) for native Americans and (Hawaiian/pacific islander) to identify Hawaiian race respectively in the context. If few people describe themselves as Latino/Hispanic with other races, like "Latino white female", in such cases, we overrode the race by Latino and put the post in the bin of Hispanic/Latino.

## 3. Age Extraction

Age is also extracted similarly to a race but in a slightly different manner. Age is extracted in the following ways: 1) Most of the post contains combined age and gender pattern like 22M/F20. Since these tokens were identified while extracting gender, so we first look for digit patterns inside the token extracted for gender. 2) If not found in step 1, then we look for patterns like digits accompanied by years/yrs/yo, etc. in the context. 3) we use patterns like "*age: digit*". 4) We look for the digits at the immediate left and immediate right position of the center word. Steps 2 and 3 are repeated for the whole post if age is not found in the context list since people may not reveal gender but may reveal age.

A summary of the self-reported demographics is depicted in column 1 and column 2 of tables 1, 2 3 for gender, race, and age respectively.

## Results

We used an accuracy score to evaluate the performance of our regex pipeline. We took a sample of random 100 posts and we were able to identify the gender with 97.6%, race with 99%, and age with a 99.7% of accuracy score. We missed a few posts where people describe their age in the format: "*I am in my mid-20s*", "*I am twenty-five*" or "*I am in my fifties*".

Table 2 presents self-reported Gender and its analysis in the posts. Out of total posts, 51% were male, 39% female, 0.79% Trans, 0.17% non-binary and 8.7% as unknown. Out of 162383 (91.3%) posts that provide a gender/sex, 56.22% were male, 42.73% were female, 0.87% were trans and 0.18% were non-binary. In posts that disclose the gender, females have the highest (51.92%) and non-binary has the lowest (43.77%) proportion of posts where at least one non- OP user commented. Females engage highest i.e., they engage in 56.6% of the female-authored posts with the commenters. From the table, it is evident that the unknown receives an average of 4.07 comments per post whereas, in the post which disclosed gender, females received the highest (2.47) average comment per post than any other gender. A similar pattern was observed for the average comment by physicians per post. Unknown received the highest (1.26) average professional comment per post. In posts that disclosed gender, females received the highest average (0.71) of professional comments per post.

Table 3 presents self-reported race/ethnicity and its analysis in the posts. Most of the post does not disclose the race, only 34254 (19.2%) of posts disclosed race. Out of the posts that disclosed race, 81.08% were written by white, 8.6% by Asian, 5.69% by Latino/Hispanic, 4.33% by black, 0.26% by American Indians, and 0.04% by Hawaiian. American Indians have the highest proportion of posts (59.55%) where at least one non-OP user commented and Hawaiian the least (23.08). Latino/Hispanic engages highest i.e., on 61.4% of the Latino/Hispanic authored posts with the commenters. From the table, it is evident that in the post which disclosed race, American Indians received the highest (2.28) average comment per post and the highest average of physicians' comments (0.8) per post than any other race.

Table 4 presents self-reported age and its analysis. Out of total posts, 163693 (92.03%) of posts provide age, 0.75% talked about Children, 20.89% about teenagers, 70.2% about adults, 5.96% about Middle age, and 2.2% about Seniors. Seniors have the highest proportion of posts (59.98%) where at least one non-OP user commented. Seniors engage highest i.e., they engaged on 64.75% of the senior-authored posts with the commenters. In the post which disclosed age, seniors received the highest average (1.05) of physicians' comments per post than any other age group. Children receive the highest overall average comment per post (3.56) among all age groups.

## Discussion

In this study, we investigated the interaction between the patients and physicians on the online platform, AskDoc. People frequently seek and receive assistance from peers in online health forums who have similar health conditions and physicians' advice without any charge. Inequalities between demographic groups in health and medical care are a long-standing and chronic problem in the US. The current study provides valuable insights by exploring the certain degree of disparity among different demographic groups on the Askdoc. We discovered that AskDoc was largely dominated by *adult white males* based on self-reported demographics. Gender and racial/ethnic minorities, however, participate in this community at lesser percentages. People seek information on online platforms to rule out the possibility of getting professional assistance offline, feeling unsure of the benefits vs. cost of clinical visits, finding a second opinion, preparing for offline medical consultation, or obtaining more details regarding the medication instructions given during visits.[20] Gender analysis shows that posts which do not disclose gender are likely to receive at least one reply from the users and if gender is revealed, females are more likely to receive at least one reply from the users as compared to other genders. This behavior can be supported by the study which says that women, on average, have bigger and more diverse social networks, with more friends and social support than any other gender.[14][15] Females receiving the highest average comment per post among all genders can also be corroborated by the finding that females engage highest with the users and physicians, hence are likely to get more comments. It is surprising to see that Unknown gender posters' engagement is the lowest (56.44%), but they have the highest both overall and physician average comments per post. Another unpredictable result that we observed was that though non-binary engages more than males, non-binary receives a low average comment per post. This finding aligns with a study by a professor at the University of Michigan who states that non-binary people face healthcare discrimination in the United States.[16]

On Race analysis, we found a smaller number of participations of all the races except whites. In the United States, Black and Hispanic adults are less likely than White individuals to own a standard computer or home broadband.[18] Black adults are more likely than White people to believe that a lack of home high-speed internet causes certain disadvantages. For example, 63% of Black and 53% of Hispanic adults compared to 49% of White adults, believe that not having high-speed internet puts them at a significant disadvantage when communicating with physicians or other medical experts.[18] Low participation by American Indians is because they are the least connected to the high-speed internet (53%) as mobile phones are the primary means by which individuals on American Indian land connect to the internet, yet many villages lack consistent mobile coverage.[19] It was surprising to see that though American Indians engage less as compared to White, Black, and Latino/Hispanic, they have the highest overall average comment per post and the highest average of physicians' comments per post among disclosed races.

Seniors engaging highest with the commentors is also one of the unexpected findings from this study. We believe this is due to the vast availability and adaptability of the internet and smartphones by the seniors. Social media use among Americans 65 and older has nearly quadrupled since 2010, while use among younger persons has stayed largely stable over the same period.[17] Children receiving the highest overall average comment per post could depict that users are more concerned about the health of children whereas seniors receiving the highest (1.05) average comment by physicians indicates physicians are more concerned about seniors' health.

The ratio of physicians to a poster in this period is 1:96 which means 96 patients for every physician. The average comment per post is 2.4 whereas the average comment by physicians per post is only 0.7 which indicates that the physicians' participation in this community is low in comparison to the posters. Some demographics received a higher response from peers and physicians as compared to the others. Completely relying on these online health forums also has its limitations. It has been discovered that some of the responses given by peers in online health groups are false and misleading.[9] As per the study by Nobles et al., physicians authored 12% of all comments as compared to 30% in our study. This shows that with time, physicians' participation is increasing. Hence, beyond the confines of clinics and hospitals, social media could bring patients and providers closer together.

## Public Health Implications

| | |
|---|---|
| Count of Unique Users | 157538 |
| Total Count of Posts | 177850 |
| Posts with at least 1 non-OP comment<br>Posts with 0 comments<br>Posts with only OP comments<br>Posts with at least one physician comment<br>Count of Unique Posters | 89050 (50.08%)<br>86450 (48.6%)<br>2350 ( 1.32%)<br>62850 (35.34%)<br>121485 |
| Total comments | 426854 |
| Total comments by physicians. | 127013 (29.8%) |
| Average comment received per post | 2.4 |
| Average comment received by physician per post | 0.71 |
| Count of Unique commenters | 85423 |
| Count of Unique physician commenters | 1641 |

Table 1: Descriptive statistics about r/AskDocs (January 2020 – May 2022)

| Gender (Posts authored by) | Total count (a) | Out of Posts that disclosed self-report gender[#] (b) | Count of posts that have at least one non-OP comment (c) | OPs engagement on posts given at least one non-OP comment (d) | Average Comment received per post | Average of Physicians' comments received per post |
|---|---|---|---|---|---|---|
| **Male** | 91284 (51.33%) | 56.22% | 43939 (48.13%) | 24868 (56.6%) | 2.07 | 0.63 |
| **Female** | 69391 (39.01%) | 42.73% | 36029 (51.92%) | 21131 (58.65%) | 2.47 | 0.71 |
| **Trans** | 1411 (0.79%) | 0.87% | 699 (49.54%) | 409 (58.51%) | 2.08 | 0.66 |
| **Non-Binary** | 297 (0.17%) | 0.18% | 130 (43.77%) | 75 (57.69%) | 1.73 | 0.54 |
| **Unknown** | 15467 (8.7%) | - | 8253 (53.36%) | 4658 (56.44%) | 4.07 | 1.26 |

Table 2: Descriptive Analysis of self-reported gender/sex
$$ : Total count of posts = 177850
\# : Posts which disclose gender = 162383

| Race | Total count (a) | Out of Posts that disclosed self-report race[#] (b) | Count of posts that have at least one non-OP comment (c) | OPs engagement on posts given at least one non-OP comment (d) | Average Comment received per post | Average of Physicians' comments received per post |
|---|---|---|---|---|---|---|
| **White** | 27773 (15.62%) | 81.08% | 13624 (49.05%) | 8165 (59.93%) | 2.23 | 0.65 |
| **Asian** | 2947 (1.66%) | 8.6% | 1485 (50.39%) | 835 (56.23%) | 1.94 | 0.66 |
| **Latino/Hispanic** | 1950 (1.1%) | 5.69% | 969 (49.69%) | 595 (61.4%) | 2.12 | 0.64 |
| **Black** | 1482 (0.83%) | 4.33% | 810 (54.66%) | 473 (58.4%) | 2.23 | 0.73 |
| **American Indian** | 89 (0.05%) | 0.26% | 53 (59.55%) | 30 (56.6%) | 2.28 | 0.8 |
| **Hawaiian** | 13 (0.01%) | 0.04% | 3 (23.08%) | 1 (33.33%) | 0.69 | 0.31 |
| **Unknown** | 143596 (80.74%) | - | 72106 (50.21%) | 41042 (56.92%) | 2.45 | 0.73 |

a% = a/$$, b% = a/#, c% = c/a, d% = d/c

Table 3: Descriptive Analysis of self-reported Race/Ethnicity
$$ : Total count of posts = 177850
\# : Posts which disclose race = 34254
a% = a/$$, b% = a/#, c% = c/a, d% = d/c

| Age Group | Total count (a) | Out of Posts that disclosed self-report age# (b) | Count of posts that have at least one non-OP Comment (c) | OPs engagement on posts given at least one non-OP comment (d) | Average Comment received per post | Average of Physicians' received comments per post |
|---|---|---|---|---|---|---|
| **Children (1-12)** | 1222 (0.69%) | 0.75% | 679 (55.56%) | 416 (61.27%) | 3.56 | 0.94 |
| **Teenager (13-19)** | 34182 (19.22%) | 20.89% | 17022 (49.8%) | 8932 (52.47%) | 2.12 | 0.58 |
| **Adult (20-39)** | 114887 (64.60%) | 70.2% | 56797 (49.44%) | 32990 (58.08%) | 2.23 | 0.66 |
| **Middle Age (40-** | 9774 (5.5%) | 5.96% | 4912 (50.26%) | 3161 (64.35%) | 2.5 | 0.77 |
| **Senior (60+)** | 3628 (2.04%) | 2.2% | 2173 (59.9%) | 1407 (64.75%) | 3.1 | 1.05 |
| **Unknown** | 14157 (7.96%) | - | 7467 (52.75%) | 4235 (56.72%) | 4.07 | 1.3 |

Table 4: Descriptive Analysis of self-reported Age
$$ : Total count of posts = 177850
\# : Posts which disclose age = 163693
a% = a/$$, b% = a/#, c% = c/a, d% = d/c